\documentclass[11pt]{article}
\usepackage{latexsym,graphicx,multirow}
\usepackage{float}
\usepackage{amssymb}
\usepackage{amsmath}
\usepackage{amscd}
\usepackage{amsthm}
\usepackage[left=2cm,top=2.5cm,right=2.5cm,bottom=1.5cm]{geometry}
\usepackage{hyperref}
\usepackage{epstopdf}
\usepackage{cite}
\usepackage{caption}
\usepackage{subcaption}
\usepackage{color}
\usepackage[utf8]{inputenc}
\usepackage[T1]{fontenc}

\begin{document}
	
\begin{center}
\large{\bf{Evaluation of Transit cosmological model in $f(R,T^{\phi})$ theory of gravity }} \\
		\vspace{10mm}
\normalsize{ Bhojraj Singh Jayas$^1$ and  Vinod Kumar Bhardwaj$^{2}$   }\\
		\vspace{5mm}
\normalsize{$^{1,2}$Department of Mathematics, GLA University, Mathura-281 406, Uttar Pradesh, India}\\
		\vspace{2mm}
$^1$E-mail:bhojraj.jayas@gla.ac.in\\
		\vspace{2mm}
$^2$E-mail: dr.vinodbhardwaj@gmail.com\\
				
		\vspace{10mm}
		
	\end{center}

\begin{abstract}
We have explored a transitioning cosmic model, depicting late-time accelerated expansion in $f(R,T^{\phi})$ theory of gravity for an isotropic and homogeneous universe, where the trace of energy-momentum tensor $T^{\phi}$ is the function of the self-interacting scalar field $\phi$. We have proposed an explicit solution to the derived model by utilizing a scale factor of the hybrid form $a(t) = t^{\alpha} e^{\beta t}$, where $\alpha$ and $\beta$ are constants. To evaluate the best-fit values of free parameters of the suggested model, the statistical analysis based on the Markov Chain Monte Carlo (MCMC) method has been employed on 57 OHD points. We have described the dynamical features of the model like energy density, cosmic pressure, and equation of state parameter in the context of scalar field $\phi$. We have also described the potential and behavior of the scalar field for quintessence and phantom scenarios. The deceleration parameter depicts a transitioning universe with signature flipping at $z_t = 0.82$ with the present value of deceleration parameter $q_0=-0.41$. The violation of SEC for the derived model indicates the cosmic expansion at a faster rate. We have used statefinders to diagnose the model. The findings for our theoretical model indicate that the derived model agrees with observed findings within a particular range of limitations.	
\end{abstract}
\section{Introduction} 
Various cosmological studies suggest that we are living in the scenario of accelerated cosmic expansion. The observational findings of $H(z)$ data, Ia Supernovae \cite{ref1,ref2}, Wilkinson Microwave Anisotropy Probe (WMAP) \cite{ref3,ref4}, Baryon Acoustic Oscillations (BAO) \cite{ref5}, Cosmic Microwave background (CMB) \cite{ref6}, Large scale structure (LSS) \cite{ref7,ref8} also support the present era of cosmic expansion. These inferences suggest that an unknown kind of energy known as dark energy (DE) appears to be stored as two thirds of the universe's considerable energy density. The dark energy is considered as a mysterious type of energy which possess a massive amount of negative pressure responsible for late time expansion of cosmos at a faster rate. The cosmological constant is considered as the simplest and most natural candidate of dark energy to describe the current accelerated expansion of cosmos. Several cosmological models have developed to describe current cosmic expansion taking $\Lambda$ as the dark energy candidate \cite{ref9,ref10,ref11,ref12}. Although the standard $\Lambda$CDM model shows nice agreement with recent cosmological observations still it faces the coincidence issue and fine-tuning problem on the theoretical ground \cite{ref13}. Apart from these issues, the $\Lambda$CDM model also suffers with $H_0$ tension. To address these issues and to describe the origin and nature of dark energy, several dynamical dark energy models including modified gravity models, models based on extra dimension, and scalar field models have been proposed in literature \cite{ref14,ref15,ref16,ref17,ref18,ref19,ref20,ref21,ref22,ref23}.\\

The quintessence scalar field $\phi$ is one of the most popular and widely recognized form of DE with EoS $\omega > -1$. It is a kind of scalar field with variable density that serves as dynamical quantity in space-time\cite{ref24}. Relying on the proportion of its potential energy (PE) and kinetic energy (KE), the quintessence can be both repulsive or attractive. The scalar field is repulsive, if PE is less than KE and causes accelerated expansion of cosmos while for the value of PE larger than KE, the scalar field attracts and universe expands at slow rate.  In other words, the expansion of the universe can be expected if the KE of scalar field is very small as compared with PE i.e. $\frac{\dot{\phi^2}}{2}<<V(\phi)$. Additionally, Chaplygin gas\cite{ref25}, k-essence \cite{ref26}, quintessence \cite{ref27}, phantom ($\omega<-1$)\cite{ref28}, quintom\cite{ref29}, and the tachyons\cite{ref30} are some other proposed dynamical dark energy models. Because of the notable qualitative resemblance with DE, the scalar field models have effectively explained the inflation in the early universe and late-time cosmic acceleration\cite{ref31}. These models involved the scalar field $\phi$ as the DE term. In these model, scalar field generates a high negative pressure with slowly reducing potential. In context of scalar field, numerous theories have been proposed to describe the evolutionary dynamics of the cosmos \cite{ref31,ref32,ref33}. The tracker’s idea was first suggested by Johri\cite{ref34}, in which universe's development can be described by a unique tracker having a potential. Many observational findings strongly justify this theory of trackers. A non-minimal relationship between dark matter and quintessence\cite{ref32,ref35,ref36} and the prospect of a scalar field development driven by a non-canonical kinetic term \cite{ref26} are two instances of the theory. The presence of the scalar field is also noticed by many fundamental theories, which motivates us to study the dynamical properties of scalar fields in cosmology. In literature, a number of cosmic models have been discussed in the distinct framework of scalar field theories \cite{ref37,ref38,ref39,ref40,ref41}.\\

Harko et al\cite{ref42}, proposed the $f(R,T^{\phi})$ theory to describe the inflationary era and present an expanding scenario of the cosmos with the aid of a scalar field.  The theory depends on the “modified action principle” and includes a scalar field $T$ representing the trace of the energy-momentum tensor and the Ricci scalar curvature $R$. Moreover, the theory presents a scalar field $\phi$ that couples to the matter fields, modifying the gravitational interaction. A system of equations governing the dynamics of scalar and gravitational fields characterize the theory. These equations can be constructed, by adjusting the action using the scalar fields ($T$ and $\phi$) and metric tensor $g^{\mu\nu}$. Although the ensuing equations are very complex and challenging to resolve analytically but in some cases, the ensuing equations can be made easier by choosing a particular form for the function $f(R,T^{\phi})$. Another key arguments in favor of its consideration is that $f(R,T^{\phi})$ gravity may explain the universe's accelerated expansion without the necessity for dark energy. Furthermore, the hypothesis may lead to gravity alterations on galactic scales and provide a different explanation for the presence of dark matter. Recently, Moraes et al\cite{ref43}, explore the cosmic model in $f(R,T^{\phi})$ theory describing different dynamical eras of the universe like inflationary era, radiation dominated era, matter dominated era, and the present accelerated expansion era of cosmos. Singh et al \cite{ref44}, described the conservation of the scalar field's energy-momentum tensor by reconstruction of the $f(R,T^{\phi})$. In addition, various cosmic models have been developed in $f(R,T^{\phi})$ gravity by taking trace of the energy–momentum tensor as the function self-interacting scalar field $\phi$\cite{ref45,ref46}.\\

In the current study, we explore the dynamical features of accelerating universe in $f(R,T^{\phi})$ gravity; where, trace of energy-momentum tensor $T^{\phi}$ is the function of self-interacting scalar field $\phi$. The manuscript is organized in the following manner. In scalar field cosmology, the model and field equations of $f(R,T^{\phi})$ are presented in Sec 2. Modeling with hybrid scale factor is mentioned in Sec 3. Observational constraints on model parameters utilizing OHD dataset are given in Sec 4. Some dynamics of the suggested model are discussed in Sec 5. In Sec 6, the energy conditions for the derived model are discussed. In Sec 7, diagnosis of the model is performed using statefinders. Concluding remark is given Sec 8.  

\section{The metric and $f(R,T^\phi)$ Gravity }
We consider the gravitational action of the $f(R,T^\phi)$ theory of gravity, in which a scalar potential $V(\phi)$ self- interacts with a minimally connected scalar fields $\phi$ \cite{ref42,ref45,ref46}.
\begin{equation}
I= \int \big[{L_\phi}+\frac{1}{2}f(R,T^{\phi}) \big]\sqrt{-g}dx^4
\end{equation}
where $f(R,T^\phi)=(R+mR^2+nT^{\phi})$ is an arbitrary function of the Ricci scalar curvature R and trace of the energy-momentum tensor $T^\phi$. The matter Lagrangian of the scalar field is given by  $L_\phi=-\big[\frac{\epsilon}{2}\phi,{\mu\phi^{'}}{\mu-V(\phi)}\big]$. We use the system of units in which $8\pi{G}=1=c$. The energy-momentum tensor of matter source is read as:
\begin{equation}
T_{\mu\nu}=-\frac{2}{\sqrt{-g}}{\frac{\delta(\sqrt{-g}L_{\phi})}{\delta{g^{\mu\nu}}}}
\end{equation}
Here $g_{\mu\nu}$ is the metric tensor. We consider the matter Lagrangian $L_{\phi}$ depend on the metric tensor $g_{\mu\nu}$ only. So, the energy-momentum tensor in eq.(2) reduced to 
\begin{equation}
T_{\mu\nu}=-2{\frac{\delta{L_{\phi}}}{\delta{g^{\mu\nu}}}}+g_{\mu\nu}L_{\phi}
\end{equation}
By varying action in Eq.(1) with respect to the metric tensor $g_{\mu\nu}$, we get the following field equation of $f(R,T^\phi)$ gravity
\begin{eqnarray}
&f_R(R,T^{\phi})R_{\mu\nu}-\frac{1}{2}g_{\mu\nu}f(R,T^{\phi})-(\nabla_{\mu}\nabla{\nu}-g_{\mu\nu}\Box)f_R(R,T^{\phi})R_{\mu\nu}\nonumber\\
&=-f_{T^{\phi}}(R,T^{\phi})(\Theta\mu\nu+T\mu\nu)+T_{\mu\nu}
\end{eqnarray}
where $f_r$ and $f_{T^\phi}$ indicate the partial derivatives of $f(R,T^\phi)$ with respect to $R$ and $T^\phi$ respectively, and $\Box$ stands for d'Alembert operator defined by $\Box$  = $g^{\mu\nu}\nabla_{\mu}\nabla{\nu}$ and $\nabla_{\mu}$ denotes the covariant derivative with respect to $g_{\mu\nu}$ associated with the symmetric Levi-Civita connection and $\Theta_{\mu\nu}$ is denoted by
\begin{equation}
\Theta_{\mu\nu}=g^{\alpha\beta}{\frac{\delta{T_{\alpha\beta}}}{\delta{g^{\mu\nu}}}}=g_{\mu\nu}L_{\phi}-2T_{\mu\nu}-2g^{\alpha\beta}{\frac{\delta^2{L_{\phi}}}{\delta{g_{\mu\nu}}\delta{g^{\alpha\beta}}}}
\end{equation}
We know that 
\begin{equation}
L_{\phi}=-(\frac{\epsilon}{2}\phi,\mu\phi^{'\mu}-V(\phi))=V(\phi)-\frac{1}{2}\epsilon{\dot{\phi}^2}
\end{equation}
where dot denote the  time derivatives. Using Eq. (6) in Eq. (5) we get,
\begin{equation}
\Theta_{\mu\nu}=g_{\mu\nu}(V(\phi)-\frac{1}{2}\epsilon{\dot{\phi}^2})-2T_{\mu\nu}
\end{equation}
From Eq. (4) and Eq. (7) we get,
\begin{eqnarray}
&f_R(R,T^{\phi})R_{\mu\nu}-\frac{1}{2}f(R,T^{\phi})g_{\mu\nu}-({\nabla_\mu}{\nabla_\nu}-g_{\mu\nu}\Box)f(R,T^{\phi})\nonumber\\
&=T_{\mu\nu}+f_{T^{\phi}}(R,T^{\phi})(T_{\mu\nu}-g_{\mu\nu}(V(\phi)-\frac{1}{2}\epsilon{\dot{\phi}^2}))
\end{eqnarray}
Contracting Eq. (8) with respect to $g^{\mu\nu}$ we get
\begin{equation}
Rf_R(R,T^{\phi})-2f(R,T^{\phi})+3\Box{f_R(R,T^{\phi})}=T^{\phi}+f_T{^\phi}(R,T^\phi)(T^{\phi}-4(V(\phi)-\frac{1}{2}\epsilon{\dot{\phi}^2})
\end{equation}
From equations (8) and (9) we may recast Einstein's general relativity equation as
\begin{equation}
R_{\mu\nu}-\frac{1}{2}Rg_{\mu\nu}=\frac{(T_{\mu\nu}+T'{\mu\nu})}{f_R(R,T^{\phi})}
\end{equation}
where\\
\begin{eqnarray*}
	&T'_{\mu\nu} = (\nabla_{\mu}\nabla_{\nu}-g_{\mu\nu}\square)f_R(R,T^{\phi})  +\frac{1}{2}g_{\mu\nu}(f(R,T^{\phi})-Rf_R(R,T^{\phi}))\\
	&+f_{T^\phi}(R,T^{\phi})(T_{\mu\nu}-g_{\mu\nu}(V(\phi)-\frac{1}{2}\epsilon{\dot{\phi}^2}))
\end{eqnarray*}
For the purpose of modeling, we consider the general form of $f(R,T^\phi)$ as given below:
\begin{equation}
f(R,T^{\phi})= -(nT^{\phi}+R+mR^2)
\end{equation}
In the context of isotropic and homogeneous $f(R,T^{\phi})$ theory, the flat FLRW metric is considered as
\begin{equation}
ds^2= dt^2-(r^2({sin}^2{\theta}d{\theta}^2+d\theta^2)+dr^2)a^2(t)
\end{equation}
where $a(t)$ is scalar factor. For the above preferred metric,\\
$R_{11}=a\ddot{a}+2\dot{a}^2$,
$R_{22}=r^{2}(a\ddot{a}+2\dot{a}^2)=r^2R_{11}$,
$R_{33}=r^2{sin}^2\theta(a\ddot{a}+2\dot{a}^2)={R_{11}r^2{sin}^2\theta}$,
$R_{44}=-\frac{3\ddot{a}}{a},$\\
we know that
\begin{equation}
R=-6(\frac{\dot{a}^2+a{\ddot{a}}}{a^2})
\end{equation}
For the scalar field $\phi$ and self-interacting potential $V(\phi)$, the energy-momentum tensor of perfect fluid is read as:
\begin{equation}
T_{\mu\nu}=\epsilon\phi_{,\mu}\phi_{,\nu}-g_{\mu\nu}[\frac{\epsilon}{2}\phi{,\sigma}\phi^{\sigma}-V(\phi)]
\end{equation}
From Eq. (14), we get the following values:\\
$T_{11}=(\frac{\epsilon}{2}\dot{\phi}^2-V(\phi))a^2,$
$T_{22}=(\frac{\epsilon}{2}\dot{\phi}^2-V(\phi))a^2r^2=T_{11}r^2,$
$T_{33}=(a\ddot{a}+2\dot{a}^2)a^2r^2{sin}^2\theta={T_{22}}{sin}^2\theta$
$T_{44}=V(\phi)+\frac{\epsilon}{2}{\dot{\phi}^2}$\\
\begin{equation}
T^\phi=4V(\phi)-\epsilon{\dot{\phi}^2}
\end{equation}
From the above Eqs. (10) - (13), and Eq. (15), the following field equations are developed in $f(R,T^\phi)$ theory \cite{ref45,ref46}. \\
\begin{equation}
2\dot{H}+3H^2-6m(26\dot{H}H^2+2H\dddot{H}+12H\ddot{H}+9\dot{H}^2)=(2n-1)V(\phi)-(n-1)\frac{\epsilon}{2}\dot{\phi}^2
\end{equation}
\begin{equation}
3H^2-18m(6\dot{H}H^2+2H\ddot{H}-\dot{H}^2)=(2n-1)V(\phi)+(n-1)\frac{\epsilon}{2}\dot{\phi}^2
\end{equation}
After simplifying above Eqs.(16) and (17), the expression for scalar field ${\dot{\phi}^2}$ and potential $V(\phi)$ can be written as:
\begin{equation}
\dot{\phi}^2=\frac{-2\dot{H}+12m(4\dot{H}H^2+3H\ddot{H}+6\dot{H}^2+\dddot{H})}{(n-1)\epsilon}
\end{equation}
\begin{equation}
V(\phi)=\frac{6H^2+2\dot{H}-12m(22\dot{H}H^2+9H\ddot{H}+3\dot{H}^2+\dddot{H})}{2(2n-1)}
\end{equation}
For the FLRW universe, the energy density $\rho_{\phi}$ and cosmic pressure $p_{\phi}$ as the functions of scalar field $\phi$ are proposed as follows\cite{ref47,ref48}.
\begin{equation}
\rho_{\phi}= \frac{1}{2}\dot{\phi}^2\epsilon+V({\phi}), 
\end{equation}
\begin{equation}
p_{\phi}=\frac{1}{2}\dot{\phi}^2 \epsilon-V({\phi})
\end{equation}
The RHS of each above equations is expressed in terms arbitrary potential $V(\phi)$ and scalar field $\phi$.\\
From Eqs. (18) - (21), the equations of energy density $\rho_{\phi}$ and pressure $p_{\phi}$ for the proposed model are expressed as
\begin{eqnarray}
&\rho_{\phi}=\frac{\big[n\dddot{H}+(6-3n)H\ddot{H}+(18-14n)\dot{H}H^2+(9n-3)\dot{H}^2\big]6m}{(n-1)(2n-1)}\nonumber\\
&+\frac{3(n-1)H^2-n\dot{H}}{(n-1)(2n-1)}
\end{eqnarray}
\begin{eqnarray}
&p_{\phi}=\frac{\big[(3n-2)\dddot{H}+(15n-12)H\ddot{H}+(30n-26)\dot{H}H^2+(15n-9)\dot{H}^2\big]6m}{(2n-1)(n-1)}\nonumber\\
&+\frac{(-3n+2)\dot{H}-3(n-1)H^2}{(2n-1) (n-1)}
\end{eqnarray}
The equation of state parameter is determined by
\begin{equation}
\omega_{\phi}=\frac{6m((30n-26)\dot{H}H^2+(15n-12)H\ddot{H}+(15n-9)\dot{H}^2+(3n-2)\dddot{H}+{(-3n+2)\dot{H}-3(n-1)H^2}}{\big[(18-14n)\dot{H}H^2+(6-3n)H\ddot{H}+(9n-3)\dot{H}^2+n\dddot{H}\big] 6m+{3(n-1)H^2-n\dot{H}}}
\end{equation}
\section{Modeling with Hybrid Cosmology }
The dynamical parameters for the proposed model are expressed as the functions of Hubble parameter $H$ as mentioned in Eqs. (22)-(24). Thus, to obtain an explicit solution of the model and to determine the dynamical behavior of the universe in the framework of $f(R,T^{\phi})$ gravity,  we need an additional constraint or parameterization of the Hubble parameter as the time function. With this motivation for the ansatz, some substantial solutions have been given in literature.  The scale factor $a(t)=exp(\Lambda t)$ with positive cosmological constant $\Lambda$, describes the De Sitter universe. Further, De Sitter and Einstein suggest the power law form $a(t)=t^{2/3}$ in flat FRW space-time.  The models of accelerated expansion of the cosmos that depict a shift from deceleration to acceleration have been used by researchers over the past thirty years. According to observational findings of Planck Collaboration\cite{ref6}, WMAP Collaboration\cite{ref3,ref49,ref50}, and type-Ia supernova\cite{ref1,ref2}, a deceleration parameter as the function of time is required to depict a flipping nature of the universe from a decelerating phase to accelerating era. In the sequence, the scale factor in hybrid form efficiently explains the necessary flipping behavior of the universe with current accelerated expansion of the cosmos. Thus, to get an obvious transitioning model of the universe, we consider a hybrid scale factor of the type $a = t^{\alpha} e^{\beta t}$; where $\alpha$ and $\beta$ are arbitrary constants. Previously, several authors has utilized ansatz of such hybrid form to explore the evolutionary behaviour of transitioning universe in different frameworks\cite{ref51,ref52,ref53,ref54,ref55,ref56}. The scale factor of such hybrid form provides a time dependent deceleration parameter in the form $q=\frac{\alpha}{(\alpha+\beta t)^2}-1$, and Hubble parameter is expressed as $H =\beta +\frac{\alpha }{t}$. The above suggested hybrid scale factor depicts a deceleration expansion ($q>0$) of cosmos for $t< \frac{\sqrt{\alpha }-\alpha }{\beta }$, while for  $t> \frac{\sqrt{\alpha }-\alpha }{\beta }$, it explains the accelerated expansion era of universe ($q<0$).    \\

Thus, for the derived model, $V(\phi)$ and $\dot{\phi^2}$ are recast as:
\begin{eqnarray}
V(\phi) &=& \bigg[6 \alpha  m \bigg(22 \alpha ^2+22 \beta ^2 t^2+\alpha  (44 \beta  t-21)-18 \beta  t+6\bigg)\nonumber\\
&+&t^2 \bigg(3 \alpha ^2+3 \beta ^2 t^2+\alpha  (6 \beta  t-1)\bigg)\bigg] /\bigg[(2 n-1) t^4\bigg]
\end{eqnarray}

\begin{eqnarray}
\dot{\phi^2} &=& \bigg[2 \alpha  \bigg(t^2-12 m \left(2 \alpha ^2+2 \beta ^2 t^2+\alpha  (4 \beta  t-6)-3 \beta  t+3\right)\bigg)\bigg] / \bigg[(n-1) t^4 \epsilon\bigg]
\end{eqnarray}
The energy density and cosmic pressure for the proposed model are expressed as:
\begin{eqnarray}
\rho (\phi) &=& \bigg[-6 \alpha  m \bigg(\alpha  (3-9 n)-2 (7 n-9) (\alpha +\beta  t)^2+6 (n-2) (\alpha +\beta  t)+6 n\bigg)\nonumber\\
&+& 3 (n-1) t^2 (\alpha +\beta  t)^2+\alpha  n t^2\bigg] / \bigg[(n-1) (2 n-1) t^4\bigg]
\end{eqnarray}
\begin{eqnarray}
p (\phi)& =& \bigg[6 \alpha  m \bigg(3 \alpha  (5 n-3)+(26-30 n) (\alpha +\beta  t)^2+6 (5 n-4) (\alpha +\beta  t)+6 (2-3 n)\bigg)\nonumber\\
&-&3 (n-1) t^2 (\alpha +\beta  t)^2+\alpha  (3 n-2) t^2\bigg] / \bigg[(n-1) (2 n-1) t^4\bigg]
\end{eqnarray}
The EoS parameter for the model is read as
\begin{eqnarray}
\omega (\phi) &=& \bigg[6 \alpha  m \left(3 \alpha  (5 n-3)+(26-30 n) (\alpha +\beta  t)^2+6 (5 n-4) (\alpha +\beta  t)+6 (2-3 n)\right)\nonumber\\
&-&3 (n-1) t^2 (\alpha +\beta  t)^2+\alpha  (3 n-2) t^2\bigg] / \bigg[3 (n-1) t^2 (\alpha +\beta  t)^2+\alpha  n t^2 \nonumber\\
&-& 6 \alpha  m \left(\alpha  (3-9 n)-2 (7 n-9) (\alpha +\beta  t)^2+6 (n-2) (\alpha +\beta  t)+6 n\right)\bigg]
\end{eqnarray}

\section{Observational constraints on model parameters}
In the context of the observational $H(z)$ dataset, we bound the model parameters $H_0$, $\alpha$, and $\beta$  within the redshift range of ${0.07}\leq{z}\leq{2.36}$ in this section. The references\cite{ref57}, contain a compilation of the observable $H(z)$ dataset. In connection with redshift, the scale factor is provided by  $ a = \frac{a_0}{1+z} = t^{\alpha}e^{\beta{t}}$; where $a_0$ represents the scale factor's current value and taken as 1 for the present study\cite{ref38}. The Hubble parameter in term of redshift $z$ can be expressed as ${H(z)} = {\frac{-1}{1+z}}\frac{dz}{dt}$. Thus, the expression for Hubble parameter can be recasts as
\begin{equation}
H(z) = \beta \bigg(1+\frac{1}{\psi(z)}\bigg)
\end{equation} 
Here, $\psi(z) = \mathcal{W}\bigg(\frac{\beta(\frac{1}{1+z})^{1/\alpha}}{\alpha}\bigg)$, and $\mathcal{W}$ stands for the  Product Logarithm function or Lambert function. 
\begin{figure}[H]
	\centering
	\includegraphics[scale=0.9]{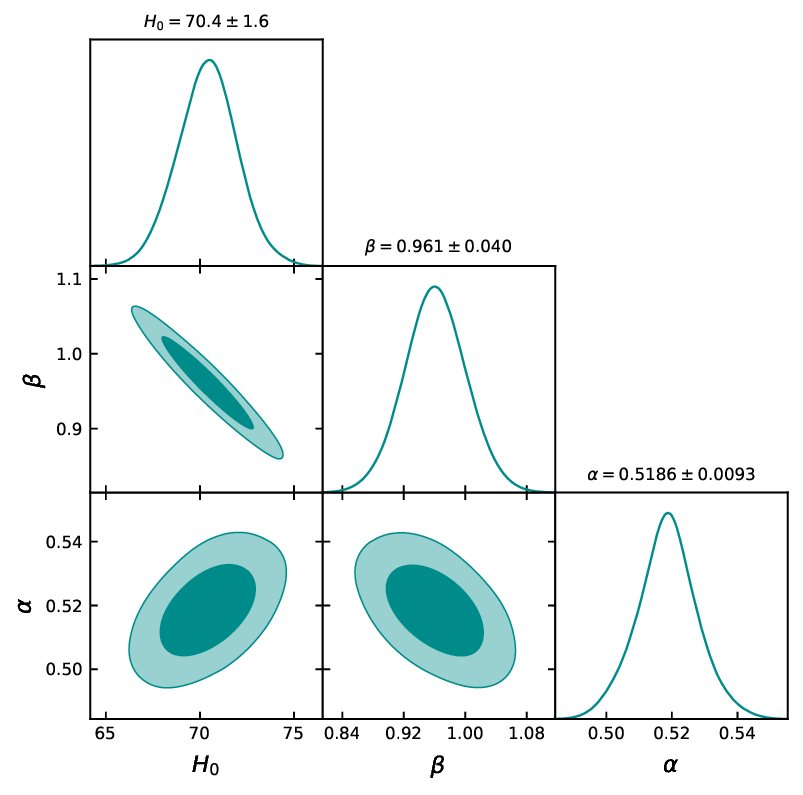}
	\caption{1-Dimentional marginal plots and 2-Dimensional contour plot with 
		68$\%$ confidence level and $95 \%$  confidence level.}
\end{figure}
We apply observational $H(z)$ data in the redshift range ${0.07}\leq{z}\leq{2.36}$ to constrain the model parameters by applying Markov chain Monte Carlo (MCMC) technique. For the purpose, freely available codes of python package EMCEE\cite{ref58} are utilized. References\cite{ref51,ref55,ref59,ref60}, provide further details on this technique. 1-Dimensional marginal plots and 2-Dimensional contour plot with 68$\%$ confidence level and $95 \%$  confidence level is shown in Figure 1. For the derived model, the optimal predicted values are estimated as $H_{0} = 70.4 \pm 1.6$, $\alpha=0.5186\pm{0.0093}$, and $\beta=0.961\pm0.040$.
\begin{figure}[H]
	(a)\includegraphics[scale=0.6]{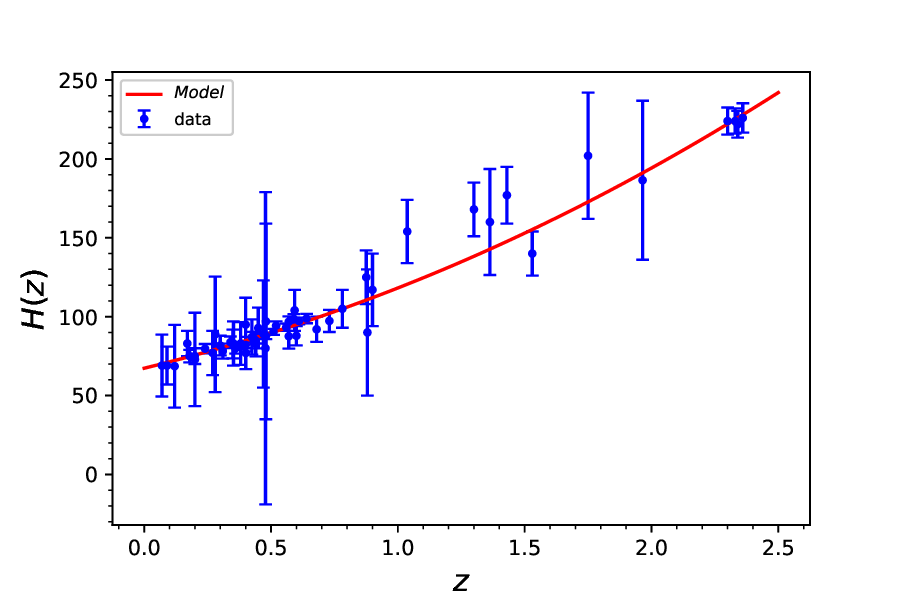}
	(b)\includegraphics[scale=0.6]{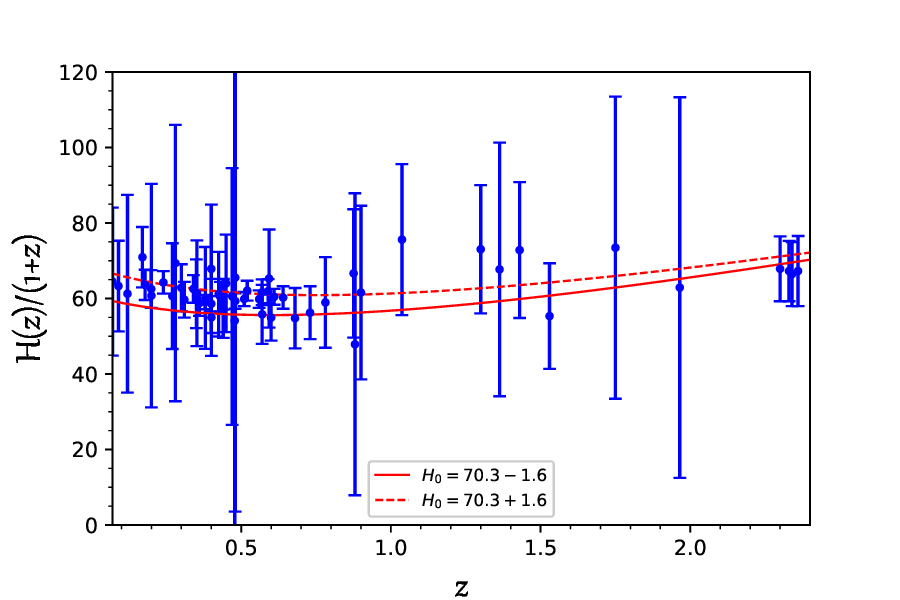}
	\caption{(a) Errorbar plot, 	(b) Plot of Hubble rate.}
\end{figure}
Figure 2, displays the best fit curves for the developed model, which plot the Hubble parameter and Hubble rate vs redshift $z$. In the figure, the points with bars indicate the Observational values of $H(z)$ data and $H_0$ is the present value of Hubble constant.
\section{Dynamics of universe}
\subsection{Deceleration Parameter}
The `deceleration parameter (DP)' $q$ is a conclusive constraint which characterizes the universe's phase shift amongst the various parameters that predict the dynamics of its evolution. From the Hubble parameter, the deceleration parameter $q$ can be calculated as $q=-\frac{\ddot{a}}{aH^2}= -1+\frac{1+z}{H(z)}\frac{dH(z)}{dz}$
\begin{figure}[H]
	\centering
	\includegraphics[scale=0.6]{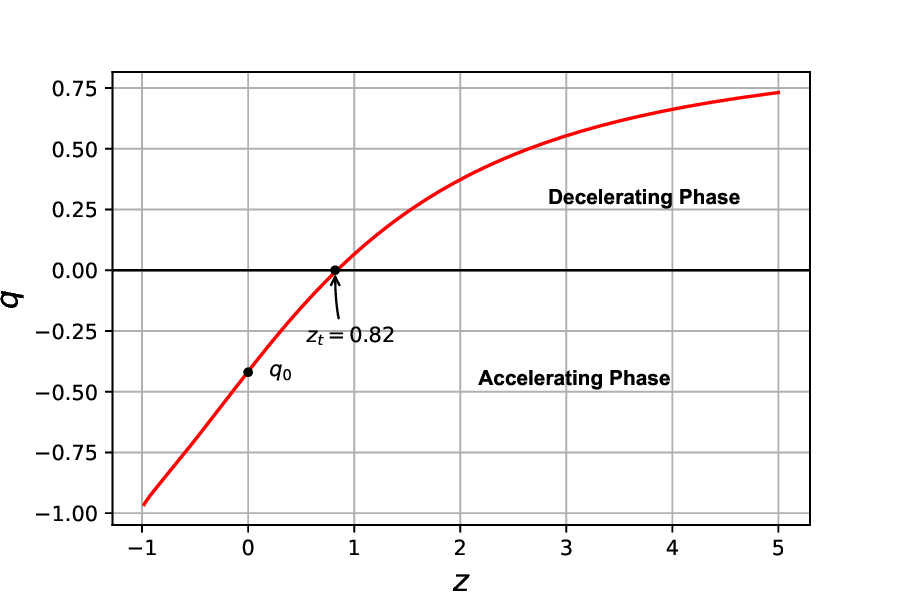}
	\caption{Plot of Deceleration parameter $q$ versus $z$ }
\end{figure}
Maximum aspects of the universe expansion are explained by the study of $q$ in conjunction with the Hubble parameter $H$. Age and the cosmic phase transitions from rapid to decelerated expansion, or the other way around, may be precisely determined from $q$ and $H$. For the suggested model, the DP is computed for the optimally predicted parameters, which are obtained by utilizing the OHD data. For the $H(z)$ dataset, Figure 3 shows the DP $q$ as the function of redshift $z$. The figure $q$ vs $z$, precisely explains the hypothesis that the earlier cosmos evolved with deceleration dynamics because of dark matter dominance. Additional analysis of the figure depicts a transitioning universe with signature flipping at $z_t = 0.82$. The present universe is expanding at a faster rate because of dark energy dominance, which will continue to be the case in the future at $z\rightarrow{-1}$ and $q\rightarrow{-1}$. This phase transition is due to an earlier slowing down of dynamics. The sign of $q$ characterizes inflation of the universe. A positive sign of $q$ i.e. $q>0$, correspond to decelerating model whereas negative sign of $q$ (particularly $-1\leq{q}<{0}$) indicates accelerating phase or inflationary model. For the derived model, the current value of DP is estimated as $q_{0} = -0.41$ approx. There is significant agreement between this value of $q_0$ and recent measurements. The suggested model's theoretically calculated results accord well with recent experimental evidence\cite{ref39,ref59,ref60,ref61,ref62,ref63,ref64,ref65}.\\

\begin{figure}[H]
	\centering
	(a) \includegraphics[scale=0.5]{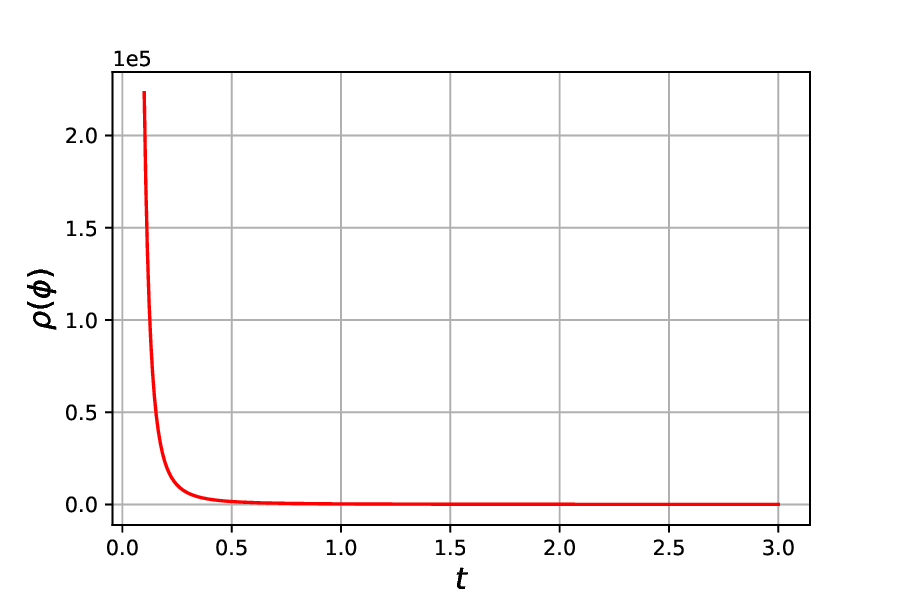}
	(b)	\includegraphics[scale=0.5]{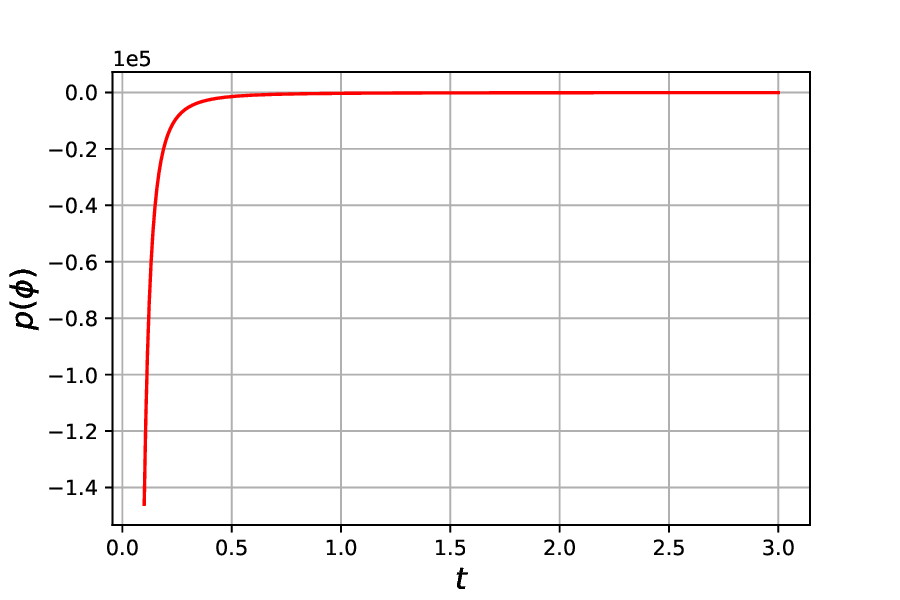}
	\caption{(a) Energy density plot vs $z$,  \, \, (b) Cosmic pressure vs $z$.}
\end{figure}
Figure 4(a), demonstrates the nature of energy density $\rho_{\phi}$ for the derived model, utilizing the best fitted values of $\alpha$ and $\beta$. The density $\rho_{\phi}$ is positive throughout the evolution of the universe. For the best fitted values of $\alpha$ and $\beta$, the nature of scalar field pressure $p_{\phi}$ is displayed in Figure 4(b). The model displays the accelerated expansion of the universe since the scalar field pressure $p_{\phi}$ remains negative throughout the evolution.\\
For best estimated $\alpha$ and $\beta$, $\omega(\phi)$ begins in the quintessence region $({-1}\textless{\omega(\phi)}\textless{0})$, and approaches to $\Lambda{CDM}$ $(\omega(\phi) =-1)$ in the late era. The model exhibits short-term fluctuations during the early epoch but soon demonstrates quintessential model features and moves closer to the $\Lambda{CDM}$. This scalar field hypothesis is confirmed by a number of cosmological measurements, such as the CMB Radiation, LSS formation, and Ia supernovae observations\cite{ref7,ref31,ref50,ref66,ref67}.
\begin{figure}[H]
	\centering
	\includegraphics[scale=0.6]{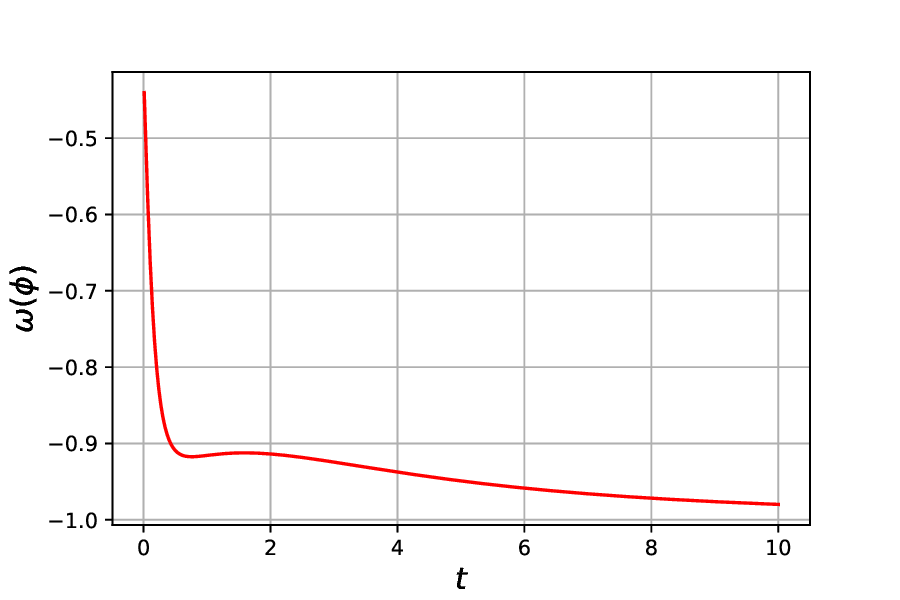}
	\caption{-Plot of EoS parameter}
\end{figure}
\subsection{ Scalar field and Potential} 
The potential is basic principle of physics which explains the energy connected to a scalar field, where each space point contains a single value\cite{ref13,ref37,ref68}. The nature of a scalar field and its potential on the particular considered physical structure. Fundamentally, the positive scalar potentials are generally linked to stable configurations since negative energies can result in unsustainable or non-material solutions. Still, there are instances in which a negative scalar potential can have physical significance, like in some cosmological inflation or DE models. In general, the distinctive nature of a scalar potential can be predicted using the fundamental physics and the precise values of the connected parameters. The behavior of a scalar potential is shown in Figure 6. During the entire evolution, the scalar potential remains positive.\\  
An equation that gives each point in space a scalar value, or just one integer, is known as a scalar field in mathematics. Classical mechanics, electromagnetic, and quantum field theory are just a few of the fields of physics that use scalar fields. Electric potential, pressure, and temperature are a few examples of scalar fields.
\begin{figure}[H]
	\centering
	\includegraphics[scale=0.6]{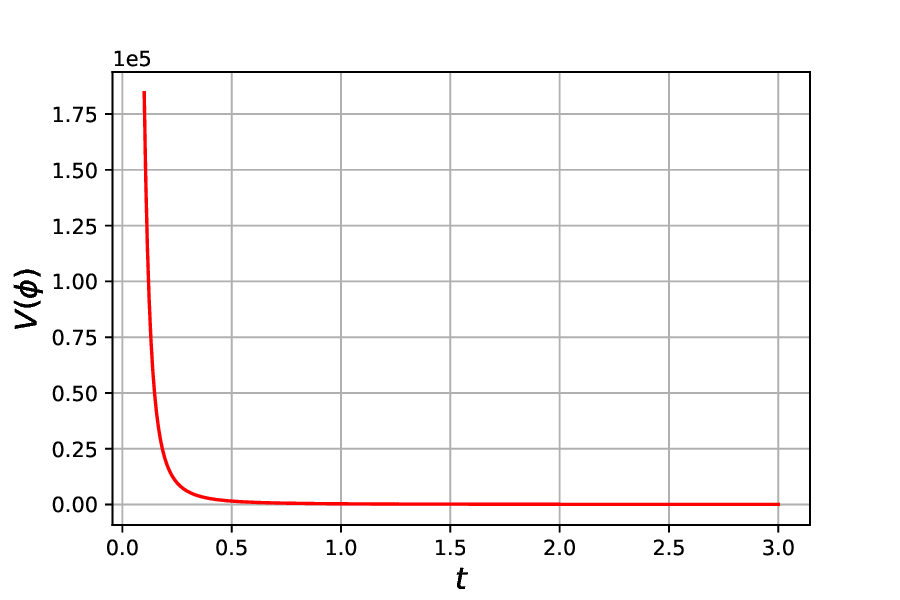}
	\caption{Plot of Scalar Potential $V(\phi)$}
\end{figure}
The scalar potential field with time is plotted in Figure 6, for best fit values of $\alpha$ and $\beta$, and for some special parameters. The figure displays that $V(\phi)$ stays positive and decreasing, and in later stages of the universe's evolution, it settles to a positive constant.
\begin{figure}[H]
	(a)\includegraphics[scale=0.5]{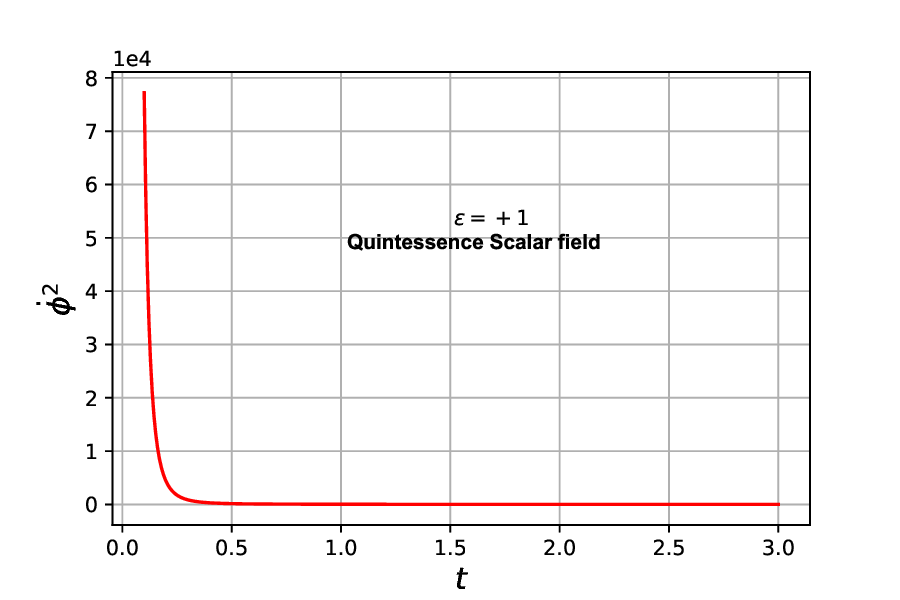}
	(b) \includegraphics[scale=0.5]{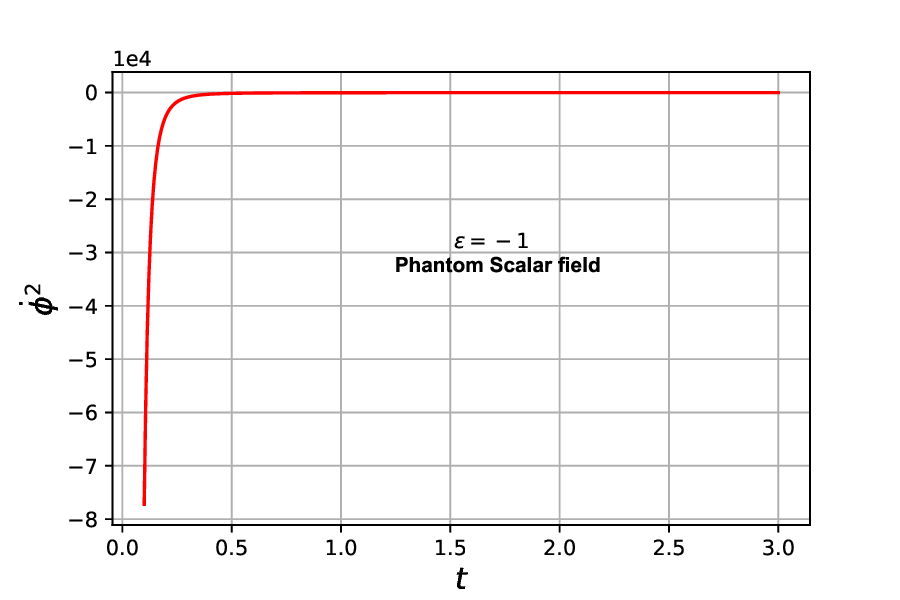}
	\caption{(a)Plot of Quintessence Scalar Field, (b)Plot of Phantom Scalar Field}
\end{figure}
{\bf 1. `Quintessence scalar Field $(\epsilon={+1})$'}\\

In case of `quintessence scalar field $(\epsilon={+1})$', from Eq. (18), we have 
\begin{equation}
\dot{\phi}^2=\frac{-2\dot{H}+12m(4\dot{H}H^2+3H\ddot{H}+6\dot{H}^2+\dddot{H})}{(n-1)}
\end{equation}
In the Quintessence scenario, the kinetic energy is positive since $\dot{\phi^2}$ is positive. In order to simulate the gravitational field and represent the current inflation of the universe, a `massless scalar field $\phi$' with a `potential $V(\phi)$' must exist \cite{ref69,ref70,ref71}. The positive behavior of quintessence scalar can be seen in Figure 7 (a).\\

{\bf 2. `Phantom Scalar Field $(\epsilon={-1})$'}\\
For the `Phantom Scalar Field $(\epsilon={-1})$', the Eq. (18) is reduced to: 
\begin{equation}
\dot{\phi}^2=\frac{-2\dot{H}+12m(4\dot{H}H^2+3H\ddot{H}+6\dot{H}^2+\dddot{H})}{-(n-1)}
\end{equation}
In the `phantom scenario model', the `scalar field $\phi$' with negative energy $\frac{1}{2}\dot{\phi}^2$ has a positive potential which is responsible for late time accelerated expansion of the universe\cite{ref14,ref71,ref72,ref73}. The figure 7 (b), shows the negative trajectory of the kinetic energy for the phantom scenario model that represents the cosmic expansion because of DE influence.
\section{Energy Condition} 
A number prevailing `energy conditions (ECs)’ in `general relativity (GR)’ that enforce limitations in order to avoid a negative energy density scenario. A collection of mathematical inequalities known as the energy conditions (ECs) which describe how matter and energy behave in space time. The universe's geodesic structure and the gravitational attraction are symbolized by these. The viability of wormhole solutions and energy conditions in modified gravity theory have examined. Refs.\cite{ref74,ref75,ref76}, provides a detailed analysis of the impact of gravity modification on some dynamical features of spherically symmetric relativistic systems. Motivated by the previously mentioned studies, we investigate the violation/validation of energy conditions and certain physical characteristics of the developed model within the context of the $f(R,T^\phi)$ gravitational theory. The energy conditions very common in `GR' can be precisely expressed as:
\begin{itemize}
	\item[i)] `WEC $\Leftrightarrow$ $\rho+p\geq{0}$ , $\rho\geq{0}$',
	\item[ii)] DEC $\Leftrightarrow$ `$\rho-p\geq{0}$',
	\item[iii)] NEC $\Leftrightarrow$ `$\rho+p\geq{0}$',
	\item[iv)] `SEC $\Leftrightarrow$ $\rho+3p\geq{0}$'.
\end{itemize}
In context of `scalar field', the energy conditions can also be expressed as: (i) `WEC $\Leftrightarrow$ $V(\phi) \geq \frac{\dot{\phi^2}}{2} $', (ii) `NEC:$\forall$ $V(\phi) \geq 0$', (iii) `SEC $\Leftrightarrow$ $V(\phi) \geq \dot{\phi^2} $', (iv) `DEC $\Leftrightarrow$ $V(\phi) \geq 0$'.
According to the DEC, every observer's measurement of the energy density must be non-negative, and any observer's measurement of the energy flux must be causal—that is, it cannot be faster than the speed of light. An observer's measurement of energy density just requires to be non-negative for the WEC, which is a less stringent version of the DEC. Every identified category of matter and energy, such as dark matter, matter particles, and electromagnetic radiation, satisfy both the DEC and the WEC \cite{ref76}. It is believed that this acceleration is being propelled by DE, which has negative-pressure energy source that dfies the SEC\cite{ref74,ref75,ref76}.

\begin{figure}[H]
	\centering
	\includegraphics[scale=0.60]{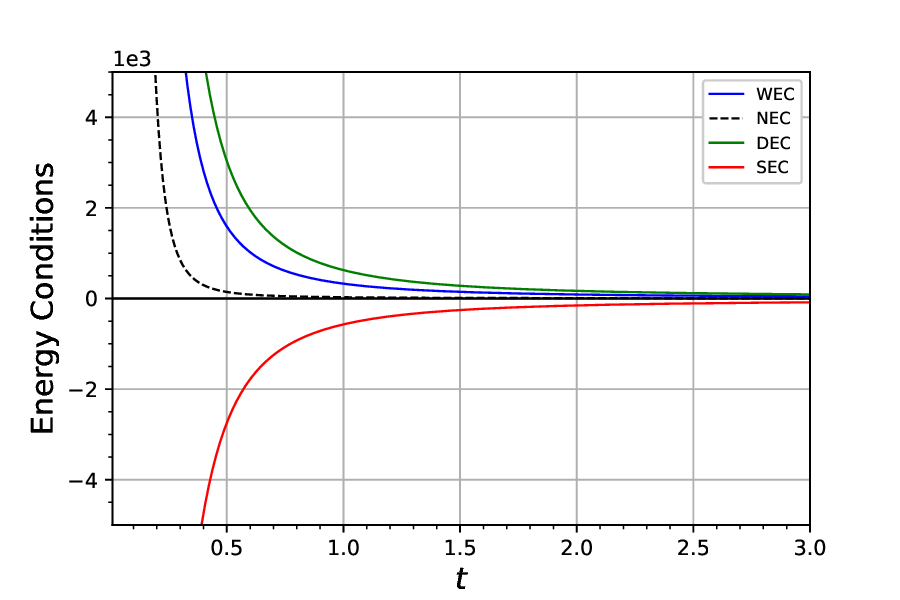}
	\caption{Plot of Energy conditions}
\end{figure}
For the suggested model, using Eqs. (27) and (28), the behavior of the energy conditions is described in Figure 8. The proposed model satisfies all energy conditions except SEC as seen in Figure 8. The violation of SEC for suggested model leads to an accelerated expansion of the cosmos\cite{ref45,ref46,ref77}. Thus, in the absence of the DE element ($\Lambda$) in the energy content of the universe, $f(R,T^\phi)$ gravity can explain the observed late-time acceleration expansion of the current universe in a logical way. For $\frac{\dot{\phi}^2}{2}<V(\phi)$, the suggested model describes inflation scenario of cosmic expansion which stipulates $\rho_{\phi}+3 p_{\phi} < 0$ \cite{ref13,ref37}.

\section{Statefinder diagnostics}
As we have seen that the deceleration parameter $q$ and the Hubble parameter $H$ alone are adequate to explain universe's evolutionary dynamics.  These parameters are categorized by scale factor and its derivatives of first and second order. Recently, Sahni et al \cite{ref78} and Alam et al \cite{ref79}, introduced a pair of new parameters ($r, s$) named as state-finders, to categorize the difference between various dark energy cosmic models. By identifying the evolutionary trajectory in the ${r-s}$ plane, the statefinder pair ($r,s$) supports in increasing the precision of model predictions. Presuming several kinds of dark energy, as studied in the literature \cite{ref79,ref80,ref81}, it is evident how the $(r-s)$ plane distinguishes the suggested cosmological model from the $\Lambda$CDM model. The statefinder parameters $r$ and $s$ can be expressed mathematically as:\\
\begin{equation}
r=\frac{\dddot{a}}{aH^3}  
\end{equation}
\begin{equation}
s=\frac{(-1+r)}{3(-\frac{1}{2}+q)}
\end{equation}

Thus, $r$ and $s$ parameters for suggested model can be recasts as:
\begin{equation}
r =\frac{\alpha ^3+\beta ^3 t^3+\alpha  \left(3 \beta ^2 t^2-3 \beta  t+2\right)+3 \alpha ^2 (\beta  t-1)}{(\alpha +\beta  t)^3}
\end{equation}
\begin{equation}
s=\frac{2 \alpha  (3 \alpha +3 \beta  t-2)}{3 (\alpha +\beta  t) \left(3 \alpha ^2+3 \beta ^2 t^2+\alpha  (6 \beta  t-2)\right)}
\end{equation}
\begin{figure}[H]
	\centering
	(a)\includegraphics[scale=0.5]{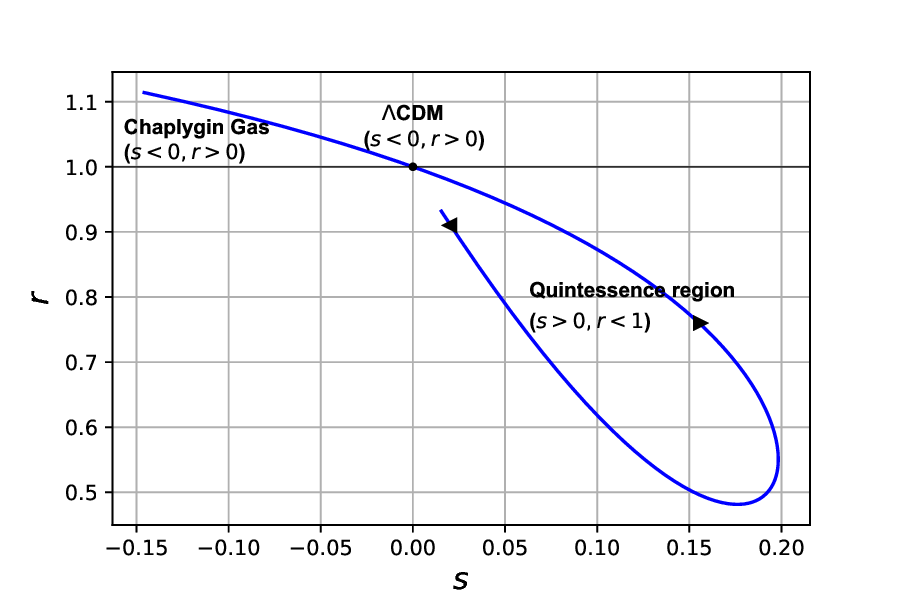}
	(b)\includegraphics[scale=0.5]{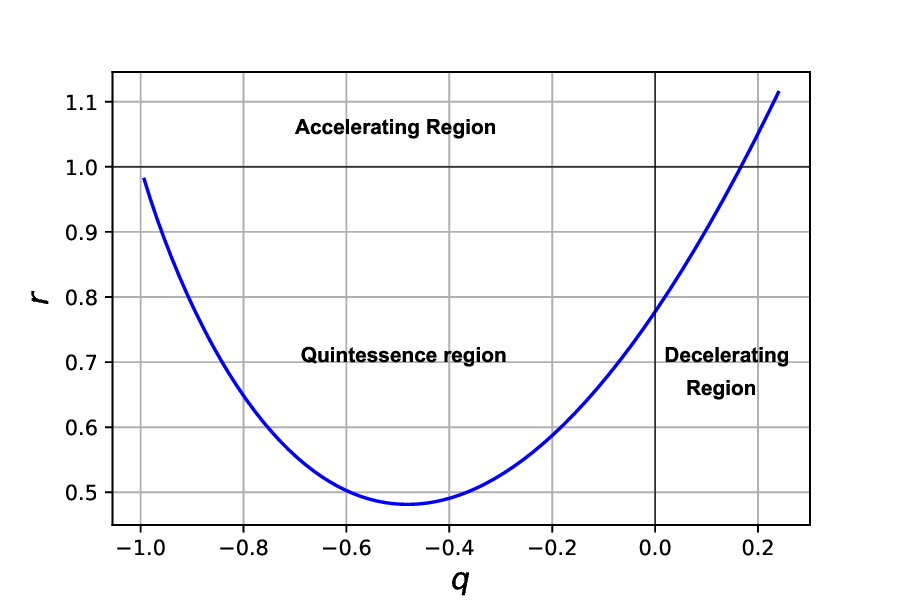}
	\caption{Plot of statefinder diagnostic}
\end{figure}
The Figure 9, illustrates the characteristics of the suggested model in the $r-s$ plane using the expressions for $r$ and $s$. Figure clearly indicates that derived model initially lies in the `quintessence era ($r<1, s>0$)' and advances to the Chaplygin gas scenario `($r>1, s<0$)' in late time after crossing $\Lambda$CDM ($r=1, s=0$). The trajectories in $r-q$ plane represents the quintessence model for $q < 0$ and $r < 1$ \cite{ref40,ref52,ref80,ref81,ref82}. The acceleration and deceleration zones of the panel are separated by a vertical line.  
\section{Concluding remarks}
In this work, we have explored a model of transitioning universe, depicting late-time accelerated expansion in $f(R,T^{\phi})$ theory of gravity. To evaluate the best fit values of free parameters of the assumed model, the statistical analysis based on Markov Chain Monte Carlo (MCMC) method have been employed on 57 OHD points. We have proposed an explicit solution to the derived model by utilizing a scale factor of the hybrid form $a(t) = t^{\alpha} e^{\beta t}$, and discussed the behavior of deceleration parameter, state-finders through graphical plotting. The main highlights of the models are:

\begin{itemize}
	\item The best fit values of model parameters are $H_0 = 70.4\pm1.6$, $\beta = 0.961 \pm 0.040$, and $\alpha = 0.5186 \pm 0.0093$. These values are estimated by statistical analysis using MCMC method and utilizing OHD dataset of 57 points. We have plotted the Hubble rate versus redshift $z$ in Figure 2, which shows a nice agreement of the derived theoretical model with $H(z)$ dataset. 
	
	\item The assumed model depicts a transitioning universe with signature flipping at $z_t = 0.82$ with the present value of deceleration parameter $q_0$ being about $-0.41$. The results found for derived model are in excellent agreement with recent observations. The proposed model represent a different dark energy models other than $\Lambda$CDM. 
	
	\item The energy density $\rho_{\phi}$ is positive and scalar field pressure $p_{\phi}$ for the derived model is negative throughout the evolution. For the best estimated values of $\alpha$ and $\beta$, the EoS parameter $\omega_{\phi}$ begins in the quintessence region $({-1}\textless{\omega_{\phi}}\textless{0})$, and approaches to $\Lambda{CDM}$ $(\omega_{\phi} =-1)$ in the late times. 
	
	\item In the Quintessence scenario, the kinetic energy is positive since $\dot{\phi^2}$ is positive. Thus, to describe the universe's current inflation, a scalar field with potential must exist that mock the gravitational field. The late time expansion of the cosmos in the phantom scenario is caused by a potential associated with a scalar field having negative kinetic energy. 
	The behavior of the model in quintessence and phantom scenarios can be seen in Figures 7 (a) \& 7 (b). 
	
	\item The trajectory in $r-s$ demonstrates that the assumed model commences in the `quintessence era' cross the `$\Lambda$CDM' and then advanced to the Chaplygin gas scenario. The trajectory in the $r-q$ plane also suggested that the proposed model is any other dark energy quintessence model different from $\Lambda$CDM. The assumed model validated all the energy conditions but violate SEC. Because of the SEC violation, a rapid expansion of the cosmos can be observed for the proposed model. 
\end{itemize}

\end{document}